\documentclass{article}

\usepackage{PRIMEarxiv}

\usepackage[utf8]{inputenc} % allow utf-8 input
\usepackage[T1]{fontenc}    % use 8-bit T1 fonts
\usepackage{hyperref}       % hyperlinks
\usepackage{url}            % simple URL typesetting
\usepackage{booktabs}       % professional-quality tables
\usepackage{amsfonts}       % blackboard math symbols
\usepackage{nicefrac}       % compact symbols for 1/2, etc.
\usepackage{microtype}      % microtypography
\usepackage{lipsum}
\usepackage{fancyhdr}       % header
\usepackage{graphicx}       % graphics
\graphicspath{{media/}}     % organize your images and other figures under media/ folder
\usepackage{geometry}
\usepackage{physics}
\usepackage{amsmath}
\usepackage{amssymb}
\usepackage{mathrsfs}
\usepackage{lineno}
\usepackage{gensymb}

\title{A Conservative Solution to the Singularity Problem in Classical GR
}

\author{
  Nikhil Bachhawat \\
  Department of Physics and Astronomy \\
  Stony Brook University \\
  Stony Brook, NY\\
  \texttt{nikhil.bachhawat@stonybrook.edu} \\
}

\begin{document}
\maketitle

\begin{abstract}
We present a conservative approach to the black hole singularity problem that remains within the framework of classical General Relativity (GR) supplemented by semiclassical quantum field theory (QFT). Our construction replaces the singular interior of a black hole with a null characteristic hypersurface that carries the exterior ADM data. The excised singular interior is replaced by a conformally flat bubble manifold. We assume the characteristic data on $\Sigma$ are shear-quiet (vanishing Bondi news and shear to leading order) so that the initial development is conformally flat (Weyl $=0$) over an early epoch. We further posit that the ingoing Hawking radiation flux provides statistically isotropic initial conditions on the bubble boundary, seeding a nearly FLRW-like epoch without first undergoing exponential inflation. The overall framework is consistent with the generalized second law of thermodynamics and offers a possible resolution of the singularity problem without invoking a full theory of quantum gravity.
\end{abstract}

% keywords can be removed

\section{Introduction}

The black hole singularity problem remains a central open issue in gravitational physics. Conventional approaches suggest that resolving singularities requires a complete theory of quantum gravity. In contrast, we present a conservative alternative: a mechanism, grounded in classical GR plus semiclassical QFT effects, that can avoid the singularity without modifying Einstein’s equations or introducing exotic matter.

The key idea is that when a black hole forms, the interior region is replaced by a null characteristic hypersurface carrying the ADM charges of the black hole in line with the membrane paradigm. This interior of the black hole is regularized by a separate space-time ``bubble" manifold, which is hypothesized to be conformally flat (vanishing Weyl curvature) and populated only with massless fields. The assumption of conformal flatness is motivated by symmetry considerations and will be treated as an ansatz for the initial epoch of the bubble.

The ingoing Hawking radiation provides a natural source of energy flux into the bubble. Due to the thermal nature of Hawking radiation. We claim that this flux can be treated as statistically isotropic on average, thereby providing initial conditions analogous to those of an FLRW-like universe without the need to postulate a separate inflationary phase. This approach provides a candidate for the resolution of the singularity problem while only relying on classical GR and semiclassical QFT results.

Similar ideas of a remnant or baby universe are considered in \cite{Hossenfelder_2010} and references therein (especially \cite{Frolov_1989,Frolov_1990,Giddings_1992, Banks_1993}). However, a mechanism for how these remnants arise is not sufficiently addressed. In this paper, we propose that a change in the Weyl invariant triggers the novel Symmetry-Restoring Phase Transition (SRPT) mechanism. This mechanism then regularizes the classical singularity at $r=0$, which allows us to probe ``after" a black hole singularity or even ``before" the Big Bang singularity.

We are able to solve information loss problems as well as provide an intuitive picture showing why the membrane paradigm works so well. We further show that our SRPT mechanism has an obvious counterpart involving a Symmetry-Breaking Phase Transition which allows the bubble manifold to transition into another epoch where the manifold resembles our initial manifold before SRPT.

The structure of the paper is as follows. In Sec. \ref{sec:SRPT_bubble}, we describe SRPT and its trigger mechanism. We discuss some of the properties of the higher-symmetry bubble manifold created by the phase transition event. We require that in-falling particles must be massless and uncharged after they pass into the bubble manifold. We assert that the bubble manifold must preserve conformal invariance symmetry and have a vanishing conformal Weyl tensor in its interior. In Sec. \ref{sec:mass_deposit}, we explore the mass deposition mechanism in SRPT and show that this leads to gravitational wave generation similar to that predicted by classical GR. Finally, in Sec. \ref{sec:penrose_diagram_structure}, we describe the large-scale structure of our updated model of the universe using a Penrose diagram.     

We emphasize that the present framework is exploratory in nature. It should be viewed as a self-consistent proposal for a curvature-triggered phase transition that replaces the singular region with a conformally flat, massless sector. The construction preserves the exterior vacuum geometry to leading order, while all quantitative aspects of the transition, such as its microphysical realization, wall dynamics, and causal development remain subjects of ongoing work.

\section{Symmetry–Restoring Phase Transition and Bubble Manifold}
\label{sec:SRPT_bubble}

\subsection{SRPT Trigger}
During generic, radiating collapse the space-time is Petrov type~I and the Weyl–spinor invariants
\begin{equation}
I \equiv \tfrac12\,\Psi_{ABCD}\Psi^{ABCD},\qquad
J \equiv \tfrac16\,\Psi_{ABCD}\Psi^{CDEF}\Psi_{EF}{}^{AB}
\end{equation}
satisfy the syzygy
\begin{equation}
\Delta \equiv I^3-27J^2 \neq 0.
\end{equation}
As the exterior settles toward a black hole geometry, it becomes algebraically special (typically type~D where $\Delta \to 0$) on the null horizon (or equivalently on the dynamical horizon in the non–spherical case). We tie the SRPT trigger to the formation of these algebraically special regions.

We now encode the matter–curvature coupling at the level of an effective field and adopt a curvature–dependent mass term,
\begin{equation}
\mathcal{L}_{\Phi}\supset -\tfrac12\,m_{\rm eff}^2\,|\Phi|^2,
\qquad
m_{\rm eff}^2 \;=\; m_0^2 - \xi_{\Delta}\,|\Delta|^{1/6},
\label{eq:meff}
\end{equation}
with \(\xi_\Delta>0\). The factor \(|\Delta|^{1/6}\) has mass dimension~2, since \([\Delta]=L^{-12}\), giving a canonically scaled correction. The exact scaling of \(\Delta\) are slicing– and solution–dependent and are deferred to later work.

Qualitatively this means that far outside the black hole \(|\Delta|\) is small and \(m_{\rm eff}^2>0\) (symmetric phase, \(\langle \Phi\rangle=0\)). As \(|\Delta|\) approaches a critical value near a null horizon, \(m_{\rm eff}^2\) turns negative in a thin band and a domain wall is nucleated. The domain wall then asymptotes to a null hypersurface \(\Sigma\) where \(\Delta\to 0\) and we have \(m_{\rm eff}^2>0\) again. Across \(\Sigma\) we posit that the ``interior" side consists of the restored higher–symmetry phase populated by massless conformally coupled fields, while the exterior remains the standard vacuum black–hole space-time.

The null characteristic hypersurface \(\Sigma\) plays a dual role. From the exterior, \(\Sigma\) carries the ADM data (mass, gauge charge, and orbital angular momentum) as surface densities or currents, analogous to the membrane paradigm. The exterior metric remains Schwarzschild/Kerr/Reissner–Nordström to leading order. From the interior, \(\Sigma\) is an initial null surface supplying characteristic data (analogous to Bondi–Sachs) for the development of a distinct space-time region, which we will refer to as the ``bubble manifold". Consistency with a vacuum exterior is enforced via null–shell Barrabès–Israel junction conditions. The full construction and stability analysis of such a shell is deferred to future work.

A slicing–independent energy balance can be expressed schematically as a flux law at \(\Sigma\):
\begin{equation}
\Delta M_{\rm ADM}^{\rm(ext)}
\;+\;
\int_{\Sigma} T_{ab}^{\text{(fields)}}\,k^{a}N^{b}\,d\mathcal{A}\,du
\;+\;
\int_{\Sigma} S_{ab}\,\nabla^{(a}k^{b)}\,d\mathcal{A}\,du
\;=\;0,
\label{eq:flux_balance}
\end{equation}
where \(k^a\) is the null generator of \(\Sigma\), \(N^a\) a transverse vector, \(S_{ab}\) the shell stress tensor, and \(T_{ab}^{\text{(fields)}}\) the bulk stress tensor. We also note that $k\!\cdot\!N=-1$

\subsection{Bubble manifold}
The SRPT leads to the formation of a domain wall which excises the black hole singularity, and instead glues \(\Sigma\) to a distinct bubble manifold. From the exterior viewpoint there is no physical cavity. \(\Sigma\) is simply the boundary of our manifold. As an ansatz for the initial epoch we hypothesize that the bubble is conformally flat (vanishing Weyl curvature) and populated only by massless fields. This conforms to the Weyl curvature hypothesis.

We model the interface $\Sigma$ as a null (characteristic) initial surface foliated by cross–sections $\mathscr{C}_u\simeq S^2$ along the generators $k^\mu=\partial_u$. To minimize free gravitational data at $\Sigma$ we choose the round conformal class on each $S^2$ and set the shear data to zero at some initial $u=u_0$. In the Bondi-Sachs formalism~\cite{Madler_2016},
\begin{equation}
N_{AB} := \partial_u C_{AB}, 
\qquad 
C_{AB}\big|_{u_0}=0,
\qquad 
N_{AB}\big|_{u_0}=0,
\label{eq:news-zero}
\end{equation}
so there is no incoming gravitational radiation at $u_0$ in an adapted frame. We adopt the following ansatz for the early bubble epoch:
\begin{equation}
g_{\mu\nu}^{\text{(bubble)}} \;=\; \Omega^2\,\eta_{\mu\nu}
\quad\text{in a neighbourhood of }\Sigma,
\qquad\Longleftrightarrow\qquad
C_{\alpha\beta\gamma\delta}=0 \ \text{(locally)}.
\label{eq:conformal-flat-ansatz}
\end{equation}
This fixes the kinematic regime used for energy/charge bookkeeping; no stronger claim is required.

Semiclassically, we adopt Unruh–like boundary conditions so that the outgoing Hawking radiation mode and late–time spectrum is unchanged. The corresponding ingoing Hawking radiation mode provides an energy inflow across \(\Sigma\) that populates the bubble. Due to the thermal nature of Hawking radiation in semiclassical treatments, we posit that the inflow is statistically isotropic, thereby seeding a near–FLRW interior without requiring an additional inflationary regime. 

Using Eq. \ref{eq:flux_balance}, we see that the exterior ADM mass obeys the slicing-independent balance law
\begin{equation}
\frac{d M_{\rm ADM}^{\rm (ext)}}{du}
=-\int_{\mathscr{C}_u} T_{ab}\,k^a N^b\, d\mathcal{A}
-\int_{\mathscr{C}_u} S_{ab}\,\nabla^{(a}k^{b)}\, d\mathcal{A},
\label{eq:flux_balance_main}
\end{equation}
where $\mathscr{C}_u \simeq S^2$ is a cross-section of $\Sigma$, and the same flux that decreases $M_{\rm ADM}$ populates the bubble through $\Sigma$.

Let $T_{uu}(u,\theta,\phi)$ denote the incoming null energy flux across $\mathscr{C}_u$. Decompose it into its spherical average and anisotropy,
\begin{equation}
T_{uu}(u,\theta,\phi)
\;=\;
\langle T_{uu}\rangle(u) \;+\; \delta T_{uu}(u,\theta,\phi),
\qquad
\langle \,\delta T_{uu}\,\rangle(u)=0,
\label{eq:flux-decomp}
\end{equation}
and define a compact anisotropy measure (collecting the $\ell\ge 2$ multipoles)
\begin{equation}
\mathcal{A}^2(u)
\;:=\;
\frac{1}{4\pi}\!\int_{S^2}\!\!\big[\delta T_{uu}(u,\theta,\phi)\big]^2\,\mathrm{d}\Omega.
\label{eq:anisotropy-norm}
\end{equation}
The spherical average $\langle T_{uu}\rangle$ changes the Bondi mass (flux balance) but, to leading order, does \emph{not} generate news:
\begin{equation}
\frac{\mathrm{d}M_{\rm ext}}{\mathrm{d}u}
\;=\;
- \!\int_{S^2}\! T_{uu}\,\mathrm{d}\Omega,
\qquad
\langle T_{uu}\rangle\neq 0 \ \Rightarrow\ N_{AB}=0 \ \text{(linear order)}.
\label{eq:mass-loss}
\end{equation}
Anisotropic flux, encoded by $\mathcal{A}(u)$, sources tidal fields and drives the shear/news, so Weyl curvature turns back on when anisotropy accumulates. At linear order in perturbations about spherical symmetry, only the $\ell\ge2$ multipoles of $T_{uu}$ source the shear/news via the Bondi constraint equations; the monopole $\langle T_{uu}\rangle$ changes the Bondi mass without generating $N_{AB}$.

When the cumulative anisotropy reaches the threshold,
\begin{equation}
\int_{u_0}^{u_*}\! \mathcal{A}(u)\,\mathrm{d}u \;\approx\; \varepsilon,
\label{eq:trigger}
\end{equation}
we allow $C_{AB}$ and $N_{AB}$ to grow and evolve the bubble away from conformal flatness. These nonzero multipole sources of anisotropy such as quantum fluctuations in Hawking modes\cite{Page_1976, Flanagan_2021}, provide the seed for symmetry breaking and transition into a different manifold with non-vanishing Weyl curvature, which resembles ours.

\section{Mass deposition mechanism} \label{sec:mass_deposit}
To ensure consistency across the null hypersurface $\Sigma$, we appeal to the Barrabès--Israel formalism for null shells~\cite{Barrabes_1991}. The jump in the transverse extrinsic curvature defines the surface stress tensor,
\begin{equation}
S_{ab} = \frac{1}{8\pi}\,[\,\mathcal{K}_{ab} - \gamma_{ab}\,\mathcal{K}\,],
\end{equation}
which can localize the conserved quantities $(M,J,Q)$ while leaving the exterior metric vacuum to leading order.

Matching of EM fields gives $4\pi\,\sigma_e=[E_\perp]_\Sigma$ and $j_A=[B_\parallel]_A$ on $\Sigma$.
For rotation, the exterior Komar angular momentum is fixed by the shell's tangential momentum density:
\begin{equation}
J=\frac{1}{16\pi}\int_{S^2} \epsilon_{abcd}\,\nabla^{c}\varphi^{d}\,,
\label{eq:KomarJ}
\end{equation}
with $\varphi^a$ the axial Killing field and the shell data are chosen so Eq. \eqref{eq:KomarJ} matches the exterior Kerr parameter.

We also introduce a macroscopic scalar field $\Phi$ representing the ADM data on a timelike ``stretched membrane’’ $\Sigma_\epsilon$ located at $r=2M+\epsilon$ (with $\epsilon>0$) in the exterior Schwarzschild geometry. On $\Sigma_\epsilon$ the induced metric is
\begin{equation}
ds^2_{\Sigma_\epsilon} = -\Big(1-\frac{2M}{r_\epsilon}\Big) dt^2 + r_\epsilon^2 \left(d\theta^2 + \sin^2\theta\, d\phi^2\right),
\label{eq:stretched_metric}
\end{equation}
and $\Phi$ satisfies a massive Klein–Gordon equation on this \emph{timelike} worldvolume,
\begin{equation}
\Box_{\Sigma_\epsilon}\,\Phi - m_{\Phi}^2\,\Phi = 0,
\label{eq:massive_KG_membrane}
\end{equation}
with physical observables defined in the limit $\epsilon\to 0^+$ and the membrane faithfully represents the physical shell at $\Sigma$.

Perturbations sourced by in-falling matter deposit energy–momentum into the membrane and, via the junction conditions, excite the gravitational quasinormal modes (QNMs) of the exterior black hole space-time. These are the ringdown modes computed by Leaver’s method and observed by LIGO/Virgo \cite{Leaver_1985,Leaver_1986_may,Leaver_1986_jul,Berti_2009,Abbott_2016,Abbott_2021_2,Isi_2019,Nair_2019}. In our picture, the dominant ringdown remains the standard GR gravitational spectrum. If a weakly coupled scalar channel $\Phi$ exists on (or near) the stretched membrane, it can in principle lead to small, model-dependent deviations (e.g., long-lived or slightly shifted modes for finite $m_\Phi$), consistent with studies of massive fields in black-hole backgrounds \cite{Konoplya_2005,Konoplya_2011,Gonzalez_2022,Addario_2024}. For $m_{\Phi}\ll 1$ (in geometric units) and small effective couplings, such deviations are expected to be subdominant to current experimental precision.

Finally, conserved quantities such as the mass, gauge charge, and orbital angular momentum (ADM data) reside on $\Sigma$ (or $\Sigma_\epsilon$ in the stretched picture) as surface densities and currents. The exterior metric remains Schwarzschild/Kerr/Reissner–Nordström to leading order. Inside the bubble manifold, post-SRPT fields are massless and carry no net ADM charges. Intrinsic helicities of massless fields may be nonzero.

\section{Large-scale structure of space-time with SRPT} \label{sec:penrose_diagram_structure}

The implications of the SRPT mechanism are apparent when considering the large scale structure of space-time in our universe and its analytical continuation. The would-be interior of a black hole is replaced by a null characteristic hypersurface $\Sigma$ that serves a dual role. From the exterior perspective, $\Sigma$ is the boundary of our manifold and carries the ADM data—mass, gauge charge, and orbital angular momentum—as surface densities and currents similar to the membrane paradigm. Hence, the exterior geometry remains a vacuum black–hole solution to leading order. From the interior perspective, the classical GR singularity has been excised and instead glued to an initial null surface supplying initial characteristic data for a distinct space-time manifold, the bubble manifold by null–shell matching.

\begin{figure}
  \centering
  \includegraphics[scale=0.55,trim=1 1 1 1,clip]{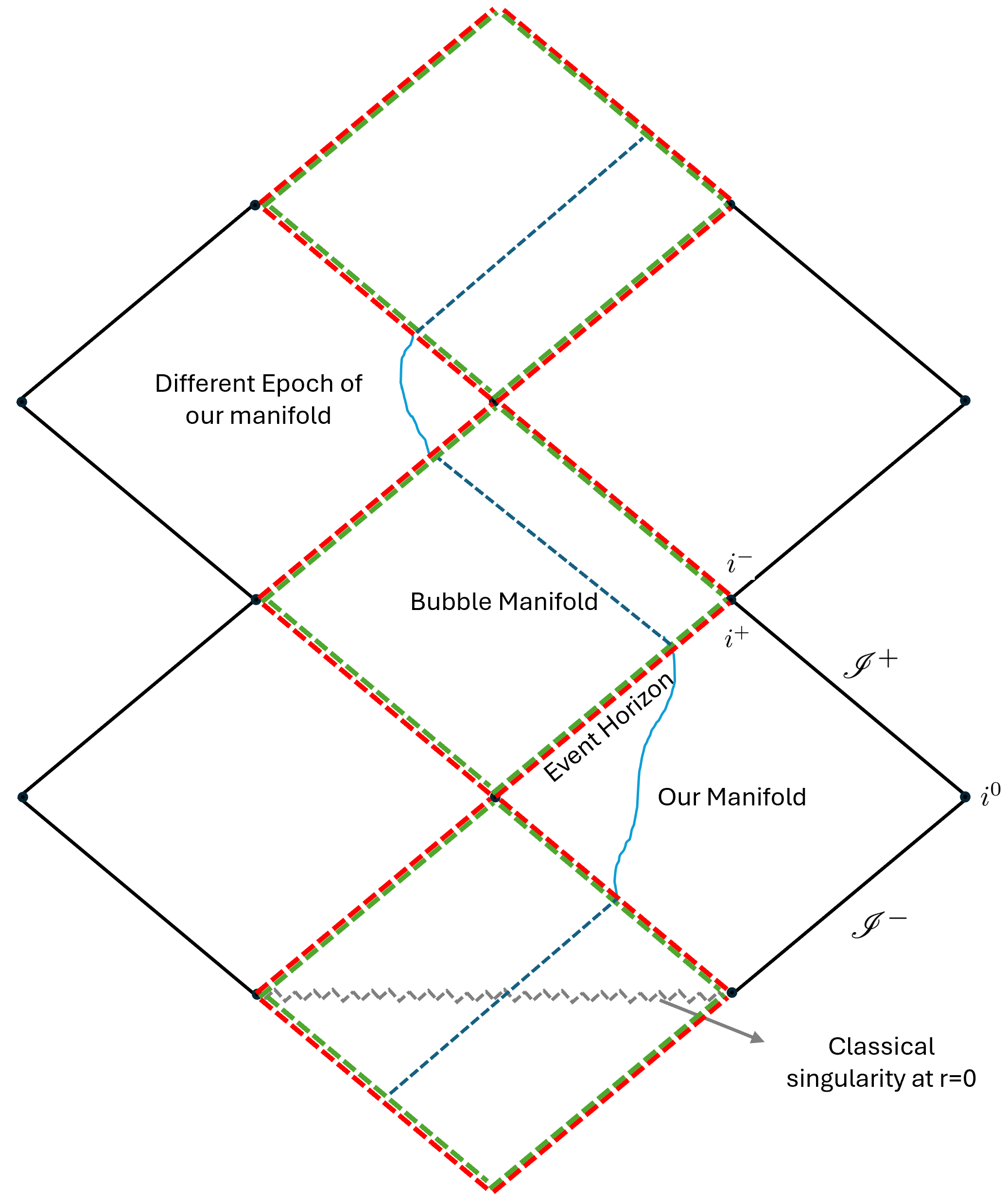}
  \caption{Penrose diagram of our universe, the bubble manifold and their extensions. The red dotted boundary around the bubble manifold is the boundary of the manifold where the geometry is the same as our manifold. The green dotted boundary is the null hypersurface with initial data for the bubble manifold and has an internal geometry with higher degree of symmetry. The blue path (including blue dotted path) represents a possible trajectory of a field/particle.}
  \label{fig:penrose_diagram}
\end{figure}

The causal picture is illustrated by the Penrose diagram in Fig. \ref{fig:penrose_diagram}. The red dotted lines represents the exterior null surface boundary, while the green dotted lines represent the initial null surface of the bubble manifold. A representative infalling worldline approaches the horizon on our side and continues as a massless trajectory inside the bubble after SRPT. We assume in our ansatz that the bubble manifold is conformally flat with vanishing Weyl tensor. This is consistent with a massless interior sector and with the Weyl–curvature hypothesis.

From a semi-classical QFT perspective, we adopt Unruh-like boundary conditions so that the outgoing (exterior) Hawking radiation modes and the late-time outgoing spectrum are unchanged. The corresponding ingoing Hawking modes contributes an energy inflow across $\Sigma$ that increases the anisotropy inside the bubble manifold. This eventually leads to a symmetry-breaking phase and a development of curvature. Due to the thermal nature of Hawking radiation, the influx of energy is statistically isotropic, providing near–FLRW initial conditions without introducing an additional inflationary regime in the early Universe. 

A covariant flux–balance relation at $\Sigma$, formulated within the Barrabès–Israel null–shell framework, relates the decrease of exterior ADM mass to the energy injected into the bubble. Considering all epochs similar to our Universe and to the bubble manifolds, we do not violate the generalized second law of thermodynamics

For any cut $\mathscr{C}_u\subset\Sigma$ of the null interface (with generators $k^\mu=\partial_u$), the generalized entropy is defined to be,
\begin{equation}
S_{\rm gen}[\mathscr{C}_u]
\;\equiv\;
\frac{\mathrm{Area}(\mathscr{C}_u)}{4G\hbar}
\;+\;
S_{\rm out}[\mathscr{C}_u],
\label{eq:Sgen-def}
\end{equation}
where $S_{\rm out}$ is the fine–grained entropy of quantum fields on the exterior side of $\mathscr{C}_u$ (including any degrees of freedom we place on a stretched interface). This follows from the quantum focusing condition~\cite{Bousso_2016,Wall_2012} under the assumed boundary conditions. This ensures that the total entropy of the exterior plus bubble region is non-decreasing along null generators of $\Sigma$,
\begin{equation}
\frac{d}{du}\,S_{\rm gen}[\mathscr{C}_u]\;\ge\;0.
\label{eq:GSL}
\end{equation}

Let $\theta$ be the null expansion on $\mathscr{C}_u$; then
\begin{equation}
\frac{d}{du}\,\frac{\mathrm{Area}}{4G\hbar}
\;=\;
\frac{1}{4G\hbar}\int_{\mathscr{C}_u}\!\theta\,\mathrm{d}A,
\qquad
\frac{d}{du}S_{\rm out}
\;=\;
\int_{\mathscr{C}_u}\! j_S(u,\theta,\phi)\,\mathrm{d}A,
\label{eq:area-entropy}
\end{equation}
with $j_S$ the (positive) entropy–flux density carried by fields crossing the cut. Hawking emission contributes to $j_S$, while focusing from energy flux $T_{uu}$ makes $\theta$ more negative via Raychaudhuri; the GSL requires that the growth of $S_{\rm out}$ compensate the area decrease.

Energy leaving the exterior black–hole sector is either radiated outward (Hawking flux) or encoded on the thin interface and into the bubble; we count the exterior radiation and any interface microstates in $S_{\rm out}$. We begin with $C_{AB}=N_{AB}=0$ (a news–free, conformally flat interior to leading order), so shear does not initially amplify the area decrease. Curvature is switched on only when anisotropic flux accumulates, at which point $S_{\rm out}$ is already increasing due to the emitted quanta. The interface does not remove entropy from the exterior ledger: any coarse–grained entropy associated with the domain–wall degrees of freedom we assign to $S_{\rm out}$ on a stretched $\Sigma_\varepsilon$.

Under these conditions,
\begin{equation}
\frac{d}{du}\,S_{\rm gen}
\;=\;
\frac{1}{4G\hbar}\int_{\mathscr{C}_u}\!\theta\,\mathrm{d}A
\;+\;
\int_{\mathscr{C}_u}\! j_S\,\mathrm{d}A
\;\ge\;0,
\end{equation}
i.e. the generalized entropy is non–decreasing along $\Sigma$ in the SRPT evolution. No extra microscopic assumptions are required beyond the flux bookkeeping and the choice to include interface entropy on the exterior side of the cut.

The global picture is compatible in spirit with CCC–like scenarios that begin in a zero Weyl curvature state, although the dynamical origin here is black–hole–triggered rather than late–aeon mass fade–out.

\section{Conclusion}

We propose a conservative mechanism, within classical GR and semiclassical QFT, that excises the singular black–hole interior, replacing it by a null characteristic boundary $\Sigma$ and a bubble manifold. This mechanism relies on a Symmetry-Restoring Phase Transition (SRPT) which is triggered by the space-time approaching algebraic speciality during black hole formation. During the SRPT process, effective mass of infalling fields/particles are suppressed by curvature–matter coupling. The ADM data remains localized on $\Sigma$, and the interior propagation proceeds as massless, conformally coupled fields on a conformally flat background. The exterior remains a vacuum black–hole solution to leading order, so standard observables such as gravitational ringdown are unchanged to leading order and any additional corrections associated with shell or bubble couplings are expected to be subleading and model dependent.

A covariant flux–balance law connects Hawking radiation induced mass loss with interior energy gain due to ingoing Hawking modes. This leads to accumulation of anisotropy which eventually leads to a symmetry-breaking phase and development of curvature. Therefore, the SRPT model provides a solution to the singularity problem of classical GR, while also providing a source for the initial data for a universe like ours. It does so without the need for an ad-hoc exponential inflationary epoch and does not violate the generalized second law of thermodynamics.

The construction presented in this paper is modest where Einstein’s equations are not modified in the exterior, and all departures from the standard picture are encoded in boundary and junction data at $\Sigma$. A full null–shell matching in the Barrabès–Israel formalism is needed to make explicit the surface stress tensor that preserves the exterior vacuum solution while localizing $M$, $Q$, and ADM $J$ on $\Sigma$. Future work should also consider stability of such a solution under perturbations more rigorously. Finally, a quantitative assessment of angular and temporal anisotropies of the ingoing flux, and of any subleading shell/bubble contributions to ringdown, would clarify the observable footprint of the mechanism.

Although we have not treated charge and rotation explicitly, the extension to Reissner–Nordström and Kerr is straightforward: surface charge and current, together with tangential momentum densities on $\Sigma$, reproduce the exterior $Q$ and $J$, while the interior remains massless and conformal. Future work will focus on deriving the explicit wall stress tensor from a Lagrangian model of the SRPT field, quantifying the timescale of curvature growth in the interior bubble, and exploring observational signatures such as subleading membrane-channel ringdown modes.

\section*{Acknowledgments}
I would like to thank George Redlinger who supported me since the initial phases of the idea presented in this paper and encouraged me to write this paper. I would also like to thank my friends (Ben, Ana and Jon) and my fianc\'ee, Leni who read my drafts and with whom I had useful conversations regarding the idea.

%Bibliography
\bibliographystyle{unsrt}  
\bibliography{references}

\end{document}